\documentclass[default,iicol]{sn-jnl}% Default with double column layout

\jyear{2023}%

\theoremstyle{thmstyleone}%
\theoremstyle{thmstyletwo}%

\theoremstyle{thmstylethree}%

\newcommand{\ramses}{{\sc ramses}}          % code names
\newcommand{\enzo}{{\sc enzo}}          % code names
\newcommand{\radmc}{{\sc radmc3d}}          % code names
\newcommand{\unit}[1]{\ensuremath{\, \mathrm{#1}}}
\newcommand{\Fig}[1]{Fig.~\ref{fig:#1}}    % Fig. reference
\newcommand{\Figure}[1]{Figure~\ref{fig:#1}}    % Figure reference

\begin{document}

\title[Rejuvenation infall]{Rejuvenating infall: a crucial yet overlooked source of mass and angular momentum}


\author*[1,2]{\fnm{Michael} \sur{Kuffmeier}}\email{mkueffmeier@mpe.mpg.de}

\author[1]{\fnm{Sigurd S.} \sur{Jensen}}\email{sigurdsj@mpe.mpg.de}

\author[3]{\fnm{Troels} \sur{Haugb{\o}lle}}\email{haugboel@nbi.ku.dk}

\affil[1]{\orgdiv{Center for Astrochemical Studies (CAS)}, \orgname{Max-Planck-Institute for Extraterrestrial Physics}, \orgaddress{\street{Gießenbachstraße 1}, \city{Munich-Garching}, \postcode{85748}, \state{Bavaria}, \country{Germany}}}

\affil*[2]{\orgdiv{Department of Astronomy}, \orgname{University of Virginia}, \orgaddress{\street{530 McCormick Road}, \city{Charlottesville}, \postcode{22902}, \state{Virginia}, \country{USA}}}

\affil[3]{\orgdiv{Center of Star and Planet Formation (Starplan), Niels Bohr Institute}, \orgname{University of Copenhagen}, \orgaddress{\street{ {\O}ster Voldgade 5-7}, \city{Copenhagen}, \postcode{1350}, \state{Sealand}, \country{Denmark}}}


\abstract{MHD models and the observation of accretion streamers confirmed that protostars can undergo late accretion events after the initial collapse phase. To provide better constraints, we study the evolution of stellar masses in MHD simulations of a $(4 \unit{pc})^3$ molecular cloud. Tracer particles allow us to accurately follow the trajectory of accreting material for all protostars and thereby constrain the accretion reservoir of the stars. 
The diversity of the accretion process implies that stars in the solar mass regime can have vastly different accretion histories. Some stars accrete most of their mass during the initial collapse phase, while others gain $>50 \%$ of their final mass from late infall. 
The angular momentum budget of stars that experience substantial late infall, so-called late accretors, is significantly higher than for stars without or with only little late accretion. As the probability of late infall increases with increasing final stellar mass, the specific angular momentum budget of higher mass stars is on average higher. 
The hypothetical centrifugal radius computed from the accreting particles at the time of formation is orders of magnitude higher than observed disk sizes, which emphasizes the importance of angular momentum transport during disk formation.
Nevertheless, we find a correlation that the centrifugal radius is highest for stars with substantial infall, which suggests that very large disks are the result of recent infall events.
There are also indications for a subtle trend of increasing centrifugal radius with increasing final stellar mass, which is in agreement with an observed marginal correlation of disk size and stellar mass.
Finally, we show that late accretors become more embedded again during late infall. As a consequence, late accretors are (apparently) rejuvenated and would be classified as Class 0 objects according to their bolometric temperature despite being $\sim 1$ Myr old.}

\keywords{star formation, protostellar environment, MHD, computational modeling}

\maketitle

\section{Introduction}\label{sec1}
Stars form in various environments of Giant Molecular clouds (GMCs). 
GMCs provide a range of physical settings such as low or high degrees of magnetization, turbulence or ionization, which shape the protostellar formation process.   
Most stars form in protostellar clusters, some form in isolation \citep{Evans2009}. 
Moreover, protostars are dynamical objects that can move through the cloud. 
They can be ejected from protostellar clusters and stars that form in isolation can become part of stellar clusters at later stages.  
Therefore, stars might travel to locations in the cloud where the conditions are different from the conditions in their birth environment. 

While (magneto-)hydrodynamic models demonstrated that variations in protostellar environments cause a diversity in the accretion process of protostars and their disks \citep{Kuffmeier+2017}, the effects of this variation are at most rudimentarily taken into account in modeling planet formation. 
Part of the reason is that planet formation was believed to only start when the star has practically accreted all of its mass and when the disk has already evolved for millions of years \citep{Lissauer1993}.  
Since the advent of ALMA, however, it has become increasingly evident that efficient dust growth already starts at an early stage when the star-disk system is still in an embedded phase \citep{ALMApartnership2015,Andrews+2018_DSHARP}. 
Theoretical studies show that early dust growth is indeed feasible \citep{Tsukamoto+2021,Bate2022,Kawasaki+2022,Lebreuilly+2023,Marchand+2023}. Therefore, widespread observations of rings, gaps and substructures already in Class I and young Class II stars have led to a paradigm shift where planet formation is assumed to start early \citep{Drazkowska+2022} fueled by a combination of small-scale instabilities, such as the streaming instability \citep{YoudinGoodmann2005,Johansen+2014}, and efficient drag-assisted pebble accretion in the gas-rich accretion disks \citep{Ormel+2010,LambrechtsJohansen2012,Popovas+2018}.
Consequentially, an early onset of planet formation implies that the processes of star and planet formation are more tightly connected than assumed in traditional models. 

While there is general consensus that ring structures in young disks are signs of early dust growth in disks, there is more controversy about their origin \citep[for a comprehensive overview of the current state of the art, see][]{Bae+2022}.
One of the most popular explanations for dust growth in disks is that planets in the disk trigger the formation of substructures including rings and gaps in the disk \citep[e.g.][]{Malik+2015,Bae+2017,Dong+2017}.
However, if rings are considered as precursors of forming planets, but planets are required to trigger rings in disks, it implies a chicken and egg problem, where the question about the formation of the first planet in the disks remains unanswered. 
To circumvent this dilemma, rings might be induced by (magneto-)hydrodynamic instabilities \citep[see][and references therein]{Bae+2022}, the back-reaction of the dust onto the gas \citep{Gonzalez+2017} (possibly enhanced by the perturbation of a (sub-)stellar fly-by \citep{Cuello+2019}), gravitational instabilities \citep{Toomre1964,VorobyovBasu2005} the presence of the (water) iceline \citep{SaitoSirono2011,RosJohansen2013}, or infall \citep{Bae+2015,Lesur+2015,Kuznetsova+2022}.
The latter scenario of infall-induced (sub-)structures is particularly interesting in the context of recent observations of "streamers" \citep{Pineda+2019,Yen+2019,Alves+2020,Garufi+2021,Huang+2021,Valdivia-Mena+2022}. Streamers are filamentary arms that connect the disk to its larger-scale protostellar environment, and were predicted in magnetohydrodynamical zoom-in simulations of star formation \citep{Kuffmeier+2017}. 

Apart from feeding and perturbing the disk directly, late infall with substantial angular momentum can cause the formation of second-generation disks \cite{Kuffmeier+2020} at large radii.
The resulting second-generation disk is likely  misaligned with respect to the already existing inner disk if the infalling material has a different orientation in angular momentum than the inner disk \citep{Thies+2011,Kuffmeier+2021}. 
While the role of infall as the trigger mechanism of origin of (sub-)structures is still debated, it has become consensus that late infall is the cause of misalignment between inner and outer disks, at least for systems with visible extended arms \citep{Ginski+2021,Rigliaco+2023}.  

An open question regarding infall is, whether it is a common feature of star formation or only an interesting, yet rare phenomenon.  
Simulations show that infall is a natural characteristic of protostellar accretion \citep{Padoan+2014,Kuffmeier+2017,Kuffmeier+2018,Kuznetsova+2019,Kuznetsova+2020}. 
Observations reveal that there are star-disk systems with clearly visible streamers that trace ongoing infall, but these systems seem to be a minority. 
(M)HD simulations, however, demonstrate that an encounter event and its associated streamers only lasts for a small fraction of a disk lifetime $\sim10^4$ to at most $10^5$ years \citep{Kuffmeier+2019,Kuffmeier+2020}. 
The intermittent nature of streamers could imply that even if streamers are only observed for a small fraction of star-disk systems, infall might nevertheless be a common feature of star formation.
A more profound understanding of the role of infall in the star formation process requires better statistical constraints.
In this article, we analyze the results from a global simulation of a star forming molecular cloud to provide such constraints on the occurrence rate and properties of infall, when taking into account for the protostellar environment. 

We structure the paper as follows. In section 2, we provide an overview of the simulations and the model setup. In section 3, we present the results of our analysis. In the first part of the section, we focus on the role of infall as a delivery mechanism of mass and angular momentum. In the second part, we examine the observational implications of infall when classifying young stellar objects (YSOs).
We discuss the results and their implications in section 4 and summarize the key results in section 5. 

\section{Methods}\label{sec2}

The simulations were carried out using the magnetohydrodynamical (MHD) version of the adaptive mesh refinement (AMR) code \ramses\ \citep{Teyssier2002,Fromang+2006}.
The setup has the same parameters as the fiducial run described in \cite{Haugbolle+2018}, but the model was carried out with a higher root grid of 512$^3$ cells (corresponding to a cell size of $\approx1600$ AU at lowest AMR level) and 6 levels of refinement with respect to the root grid (corresponding to a cell size of $\approx25$ AU at highest resolution).
This leads to effectively 8 times the number of cells. The most important criterion for refinement is density.
Output files were written with a higher cadence of 5000 years.
The simulations were previously used for a study of water deuteration by post-processing the chemical evolution of the gas carrying out radiative transfer simulations \citep{Jensen+2021}.
The simulation domain is a cubical box with periodic boundary conditions. The side length is $L_{\rm box}=4 \unit{pc}$, the total mass of the box is $M_{\rm box}=3000$ M$_{\odot}$. 
We assume the gas to be isothermal with a temperature of 10 K. Starting from a uniform distribution of density and magnetic fields, turbulence is initialized by running the simulation without self-gravity for $\sim20$ dynamical times applying random acceleration with power at wavenumbers $k$ in the range of 1 to 2, where $k=1$ corresponds to the size of the box. 
The amplitude of the driving is such that the sonic Mach number is $\mathcal{M}_{\rm s}\equiv \sigma_{\rm v}/c_{\rm s}=10$, where $\sigma_{\rm v}$ is the three-dimensional rms velocity and $c_{\rm s}$ is the isothermal
speed of sound.
The Alfv{\'e}nic Mach number is $\mathcal{M}_{\rm A}\equiv \sigma_{\rm v}/v_{\rm A}$=5, where $v_{\rm A}$ is the average Alfv{\'e}n speed $v_{\rm A}=B/\sqrt{4\pi \rho}$ corresponding to a mean magnetic field strength $B=7.2\, \mu G$. 
The virial parameter $\alpha_{\rm vir}= 5 \sigma_{\rm v}^2 R/ (3 G M_{\rm box} ) = 0.83$, where $G$ is the gravitational constant and $R=L_{\rm box}/2$ is chosen as the characteristic dynamical length scale in the box.

In addition to the public version of \ramses, the code includes sink particles and tracer particles. 
Sink particles are used as a proxy for stars to record mass accretion onto protostars. In this paper we will denominate them stars or star particles.
They are created and accrete mass according to the recipe described in detail in section 2 of \cite{Haugbolle+2018}.
The resulting initial stellar mass function obtained by the end of the simulation is in very good agreement with observations.
At the beginning of the simulation, we inject particles that are passively advected following the streamlines of the gas.
These so-called (Lagrangian) tracer particles do not affect the gas dynamics or the gravity in the simulation.
Tracer particles are a powerful tool that allow us to reconstruct the trajectories and history of the mass flow before accretion onto a star. 
Tracer particles are initialized when gravity is turned on, and the mass is calculated as the corresponding cell mass in the root grid cell, where they are created.
The total number of tracer particles is 134 million corresponding to one tracer particle per root grid cell, and hence each tracer particle correspond to a specific mass element in the box.
We investigate the role of infall in the accretion process of stars from the formation of the first star until about 2 million years later.

\begin{figure*}
\centering
\includegraphics[width=\linewidth]{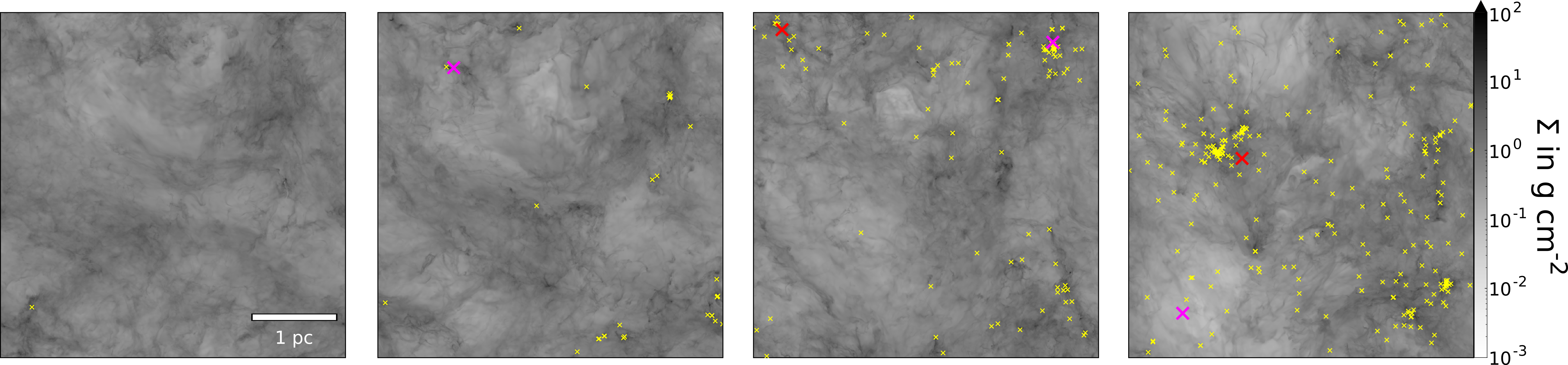}
\caption{Evolution of column density in the box as seen along the z-axis of the coordinate system. From left to right: $t \approx 2$ kyr (first snapshot after formation of the first star in the simulation), $t \approx 0.442$ Myr (first snapshot after formation of star A selected for follow-up analysis [47th star in the simulation]), $t \approx 1.022$ Myr (first snapshot after formation of star B selected for follow-up analysis [124th star in the simulation]) and $t\approx 1.995$ Myr (last snapshot of the simulation). 
Star A is highlighted in violet color, star B in red color.
Yellow crosses show the location of all other stars in the simulation at the corresponding snapshots. }\label{fig:column4pc_evo}
\end{figure*}

\section{Results}\label{sec3}
We first present the global evolution of the stars for a time span of $\approx2$ Myr as illustrated in \Fig{column4pc_evo} and we highlight the effect of infall by focusing on the mass accretion process of two individual stars. 
The stars that we label as star A and star B evolve to similar final masses ($\approx0.74$ M$_{\odot}$ and $\approx0.92$ M$_{\odot}$) by the end of the simulation. 
The locations of the selected stars at their formation times and at the end of the simulation are highlighted in \Fig{column4pc_evo}. 
The stars are representatives for the possible deviations in the accretion histories of stars, as one of the stars accretes practically all of its mass during the initial collapse phase, while the other star experiences a substantial episode of post-collapse infall. 
Using tracer particles, we also investigate the angular momentum budget of the accreting material of all the stars.
We highlight the differences in the delivery of angular momentum by comparing the angular momentum of the accretion reservoir of the two selected stars in more detail.
Finally, we present the results of continuum radiative transfer post-processing (carried out with \radmc \cite{Dullemond2012}) of the selected star that undergoes substantial late accretion. 
The radiative transfer post-processing allow us to derive the evolution of the bolometric temperature $T_{\rm bol}$, which is a common diagnostic quantity to define the evolutionary stage of a protostar.

\subsection{Infall throughout the evolution of the clump}\label{late_all}
In total, 321 stars form at various locations in the box during this period. By the end of the simulation stellar clusters establish as shown in \Fig{column4pc_evo}. 
As we are interested in answering the question how (late) infall influences the mass accretion process, we constrain the analysis in this subsection to the 104 stars that reached an age of at least 1.2 Myr by the end of the simulation.
Many stars have evolved for longer periods, while more than 200 stars have not reached an age of 1.2 Myr yet. 
We choose this age cut to include at least 100 stars in our statistical analysis, while ensuring that the selected stars have evolved well beyond the characteristic collapse period of their progenitor cores. 
The choice to compare the evolution of stars with an age of at least  1.2 Myr is the result of this compromise. 

\begin{figure*}
\centering
\includegraphics[width=\linewidth]{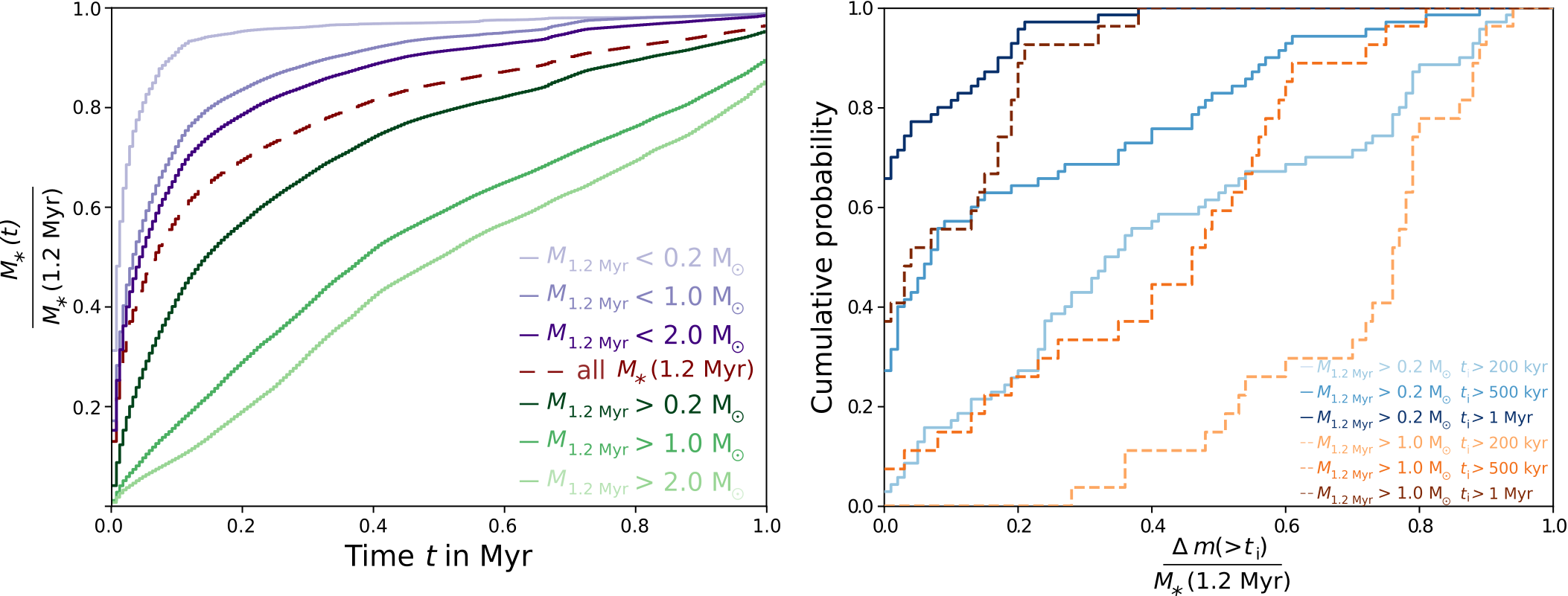}
\caption{Accretion and probability of late infall. Left: Average stellar mass evolution relative to their final mass. The average is computed by only taking into account for stars that have an age of more than 1 Myr at the end of the simulation. The dark red dashed line shows the average growth for all stars in the simulation. The light to dark purple lines mark the average relative mass growth for stars with final masses \textit{less} than $0.2 $M$_{\odot}$, $1 $M$_{\odot}$ and $2 $M$_{\odot}$. Dark to light green lines mark the average relative mass growth for stars with final masses \textit{greater} than $0.2 $M$_{\odot}$, $1 $M$_{\odot}$ and $2 $M$_{\odot}$.  
Right: cumulative probability distribution of late infall relative to the final mass of the star. The x-axis indicates how much mass with respect to the final mass accretes onto the considered stars after time $t_{\rm i}$ on average. The darkness of the colors corresponds to increasing time thresholds of $t_{\rm i} = 200$ kyr, $t_{\rm i} = 500$ kyr and $t_{\rm i} = 1$ Myr). The blue solid (orange dashed) curves show the distribution for all stars with a final mass greater than 0.2 M$_{\odot}$ (1 M$_{\odot}$).}\label{fig:sink_mass}
\end{figure*}

In \Fig{sink_mass}, we show the average evolution of stellar mass (left panel) for these 104 stars. 
The average mass evolution is plotted relative to the stellar mass $M_{1.2\rm Myr} = M_{*}(1.2\ \rm Myr)$ at $t=1.2$ Myr.
Apart from the average stellar mass evolution of all the 104 stars that have reached an age of at least 1.2 Myr, we also compare the average evolution of these stars with masses $M_{1.2\rm Myr}$ below/above thresholds of 0.2 M$_{\odot}$, 1 M$_{\odot}$ and 2 M$_{\odot}$ (see Table \ref{sink_mthresh} for an overview of the number of stars above/below the selected thresholds). 
\begin{table}[th]
\centering
\begin{tabular}{ccc}
     $M_{\rm thr}$ in M$_{\odot}$ & $N(M_{*}\leq M_{\rm thr})$ & $N(M_{*}>M_{\rm thr})$ \\ \hline
0.2 & 34 & 70   \\
1.0 & 75 & 29   \\
2.0 & 88 & 16  \\

\end{tabular}
\caption{Number of stars with ages of at least $1.2$ Myr by the end of the simulations above/below the selected mass thresholds chosen for the mass evolution plot in \Fig{sink_mass}. The left column shows the mass thresholds of 0.2 M$_{\odot}$, 1 M$_{\odot}$ and 2 M$_{\odot}$, the middle column shows the number $N$ of stars with masses below the threshold and the right column shows the number of stars above the threshold.}
\label{sink_mthresh}
\end{table}

The comparison shows a clear trend.
The higher $M_{1.2\rm Myr}$, the bigger the contribution of late accreting material. 
While the stars with low masses at 1.2 Myr of $M_{1.2\rm Myr} < 0.2$ M$\_{\odot}$ accrete almost all of their mass during the first $\approx 100$ kyr and thereby during the initial collapse phase, the stars with higher $M_{1.2\rm Myr}$ gain on average more of their mass at later stages.
In fact, stars with masses $M_{1.2\rm Myr} > 1$ M$_{\odot}$ accrete about 50 $\%$ of their final mass 500 kyr after their birth and $20 \%$ after 1 Myr. 

Although not shown in the plot, some of the more massive stars that have evolved for longer time than 1.2 Myr by the end of the simulation continue to accrete substantial amounts of mass at even later stages.
This is in good agreement with the inertial flow model \citep{Padoan+2020}, as well as results found by \cite{Smith+2011} and \cite{Pelkonen+2021} that the fraction of accreting material that is initially unbound increases with increasing final stellar mass.

The left panel shows a clear trend, but the average does not reflect the more chaotic nature and diversity of the accretion process associated with heterogeneous accretion.
The right panel in \Fig{sink_mass} shows the cumulative distribution of the relative amount of mass that accretes onto a star after 200 kyr, 500 kyr and 1 Myr after their formation to the stellar mass at 1.2 Myr. 
This plot provides a better overview of the diversity of accretion histories among the stars, while also showing the same trend of increasing probability of late infall for stars with larger final masses. 
It shows for instance that for stars heavier than $0.2 \unit{M}_{\odot}$, about $20 \%$ accrete more than $50 \%$ of their mass at $1.2 \unit{Myr}$ after the first 500 kyr. 
The fraction of stars experiencing substantial late infall is higher for stars with higher $M_{1.2\rm Myr}$.
For stars heavier than 1 M$_{\odot}$, more than $40 \%$ accrete $>10 \%$ of their mass at $1.2$ Myr after 1 Myr.
$80 \%$ of the stars with $M_{1.2\rm Myr}>1 \unit{M}_{\odot}$ accumulate more than 50 $\%$ of $M_{1.2\rm Myr}$ after 200 kyr. 
40 $\%$ of the stars with $M_{1.2\rm Myr}>1 \unit{M}_{\odot}$ have not even accreted 50 $\%$ of $M_{1.2\rm Myr}$ within the first $0.5$ Myr.
We also point out that the contribution of late infall would yet be higher if we had chosen a later reference point in time than $1.2$ Myr to compare the stellar masses in this plot.

\subsection{Comparison of the individual mass accretion process for two stars with and without infall}\label{late_ind}
\begin{figure*}
\centering
\includegraphics[width=\linewidth]{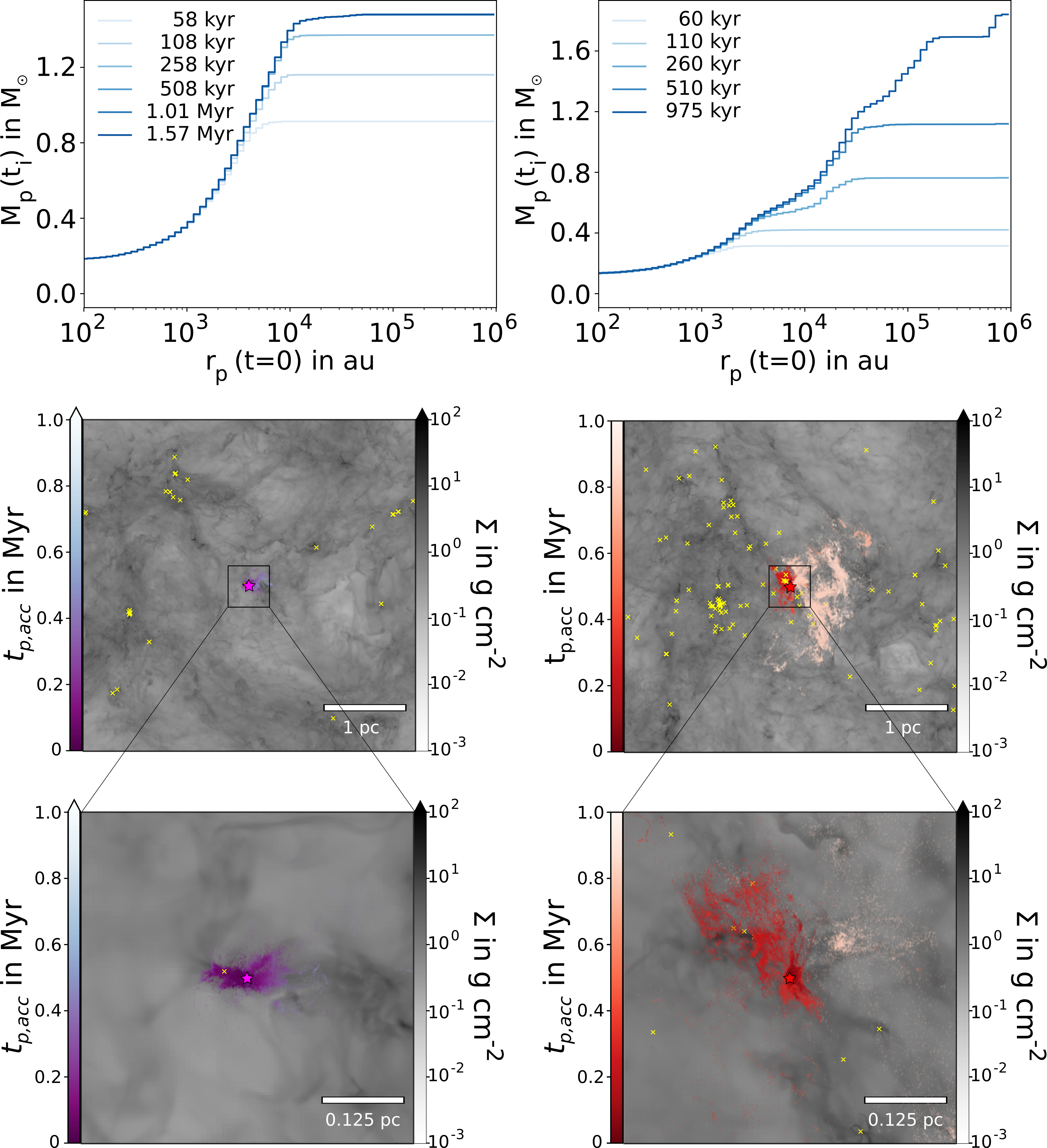}
\caption{Top panels: Two examples illustrating the diversity in accretion histories for two stars that accrete to about the same final mass. Left: star A that accretes almost all of its final mass during the initial collapse phase. Right: star B that accretes most of its final mass after the initial collapse phase.
The x-axis displays the radial distance of the accreting tracer particles from the protostar $r_{\rm p}(t)$ at $t=0$. The y-axis displays the total amount of mass $M_{\rm p}(t_{\rm i})$ that accretes onto the protostar by time $t_{\rm i}$. The stellar mass $M_*(t)$ at $t=t_{\rm i}$ is $0.5 M_{\rm p}(t_{\rm i})$ because of the efficiency parameter in the accretion recipe.  
The solid lines show how far away from the star the gas that has accreted by time $t_{\rm i}$ was at the time of star formation ($t_{*}=0$). 
The darkness of the lines from light to dark blue indicates increasing the increase of $t_{\rm i}$ ($t_{\rm i} \approx 60$ kyr, 110 kyr, 260 kyr, 500 kyr, 1 Myr (only star A) and the last snapshot of the simulation.  
Middle and bottom panels: Column density at time of formation of star A (left) and star B (right) within a region of $\pm 2$ pc (middle) and $\pm 0.25$ pc (bottom) around the stars. The dots in the panels represent the gas that accreted onto the star by the end of the simulation. The darkness of the color of the tracer particles indicates when they accreted onto the central star. The later the gas accretes the lighter the color in which they are marked.)
}\label{fig:tracer_cum_46_123}
\end{figure*}

In \Fig{tracer_cum_46_123} we show the cumulative accumulation of tracer particles for two representative cases for protostars with and without substantial late infall. By the end of the simulation,  both protostars have reached a similar final mass. Star A has a mass of $M_{\rm star A}\approx0.74$ M$_{\odot}$ and star B of $M_{\rm star B}\approx0.92$ M$_{\odot}$ at the end of the simulation. 
Given the efficiency parameter for sink accretion of 0.5, this corresponds to a total mass of accreting tracer particles of $M_{\rm p, star A}\approx1.48$ M$_{\odot}$ and star B of $M_{\rm p, star B}\approx1.84$ M$_{\odot}$. As shown in \Fig{tracer_cum_46_123}, the stars are accreting their mass in very different ways. 
While accretion onto star A has practically stalled after the initial collapse phase, star B undergoes substantial accretion at a later stage. 
Star A accretes almost all of its final mass during the initial collapse phase, while star B accretes more than $1/3$ of its mass after more than 500 kyr.
The plots illustrate the implications that protostars rarely form in isolation and can also travel through the interstellar medium after their formation (see \Fig{column4pc_evo}).  Star B travels through the clump, enters and leaves a region of a stellar cluster towards the end of the simulation.  Even star A that has an accretion history that is in agreement with a traditional picture of accretion from a relatively narrow mass reservoir forms in fact as part of a binary system. A deeper analysis of the role of multiplicity in the star formation process is beyond the scope of this paper, but we refer the reader to \cite{KuruwitaHaugboelle2022} for a detailed analysis of the formation and evolution of systems of order binary and higher. 

\Fig{tracer_cum_46_123} illustrates that the accreting material of star B was initially $\sim10^5$ AU away and thereby not bound to the progenitor core whose collapse originally led to the formation of the protostar.
The material that is accreting late is hence unbound from the progenitor core and is captured by the protostar during its journey through the cloud \citep[see][for a detailed analysis of the properties of the progenitor core and its relation to the final stellar mass]{Pelkonen+2021}. 

\subsection{Angular momentum budget and the role of infall as a delivery mechanism}
Disks form as a result of conservation of angular momentum. It is generally assumed that disks form during the initial collapse phase and evolve as isolated entities afterwards. 
\Figure{sink_mass} demonstrates however that many stars are fed with substantial amount of material that stems from reservoirs far away from the progenitor core. \cite{Kuffmeier+2020} showed that encounter events can in principle lead to a second generation of disks if the infalling material has high enough angular momentum. We do not resolve disks in our simulations and transport of angular momentum from the disk properly, but we can measure the amount of angular momentum of the infalling gas before it falls onto the system. 

\begin{figure*}
\centering
\includegraphics[width=\linewidth]{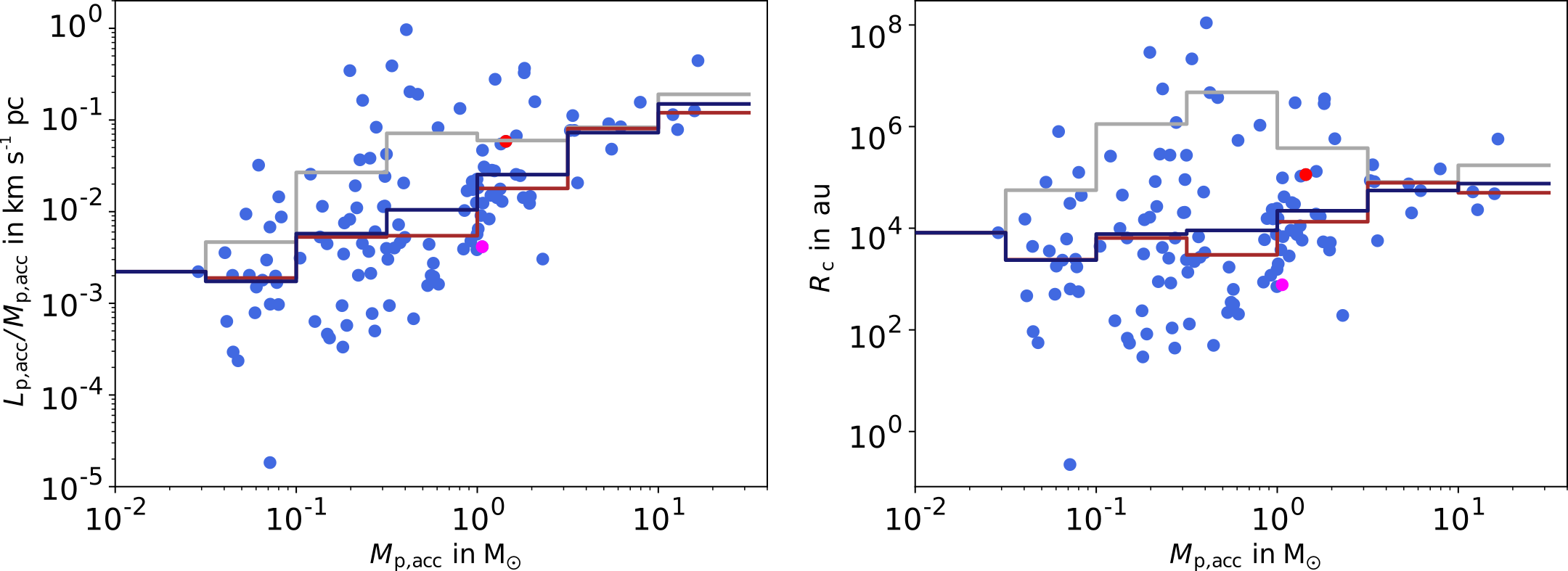}
\caption{Specific angular momentum (left panel) and centrifugal radius $R_{\rm c}$ (right panel) of accreting particles at the first snapshot after star formation over stellar mass at the end of the simulation. The specific angular momentum is computed from all particles that have not accreted onto the star by the first snapshot after star formation, but have accreted onto the protostar by the end of the run. The plot only shows the formation of 124 stars that have evolved for at least 974 kyr by the end of the simulation. The dots in violet color corresponds to the profile for star A, the dots colored in red to the profile for star B. The grey line shows the arithmetic mean per logarithmically-spaced mass bin, the dark blue line shows the geometric mean and the brown line shows the median per bin. The bin size is such that there are two bins per magnitude 10. Note that the total mass of accreting particles is twice the final stellar mass because of the choice of 0.5 for the accretion efficiency parameter to account for the mass loss in outflows.}\label{fig:M_J_M_Rc}
\end{figure*}

\subsubsection{Distribution of specific angular momentum for stars with and without infall}
In the left panel of \Fig{M_J_M_Rc}, we show the specific angular momentum of all accreting tracer particles at the time of star formation for all of the 124 stars in the simulation that have evolved for at least 975 kyr. (We choose this age threshold to include star B in the selection.) 
The specific angular momentum is 
\begin{equation}
J=L/M,    
\end{equation}
where 
\begin{equation}
    L = \lvert \Sigma_{i=1}^N m_{\rm p,i}  \mathbf{r}_{\rm p,i} \times \mathbf{v}_{\rm p,i} \rvert
\end{equation} 
is the total angular momentum of all $N$ accreting particles with individual particle mass $m_{\rm p,i}$, relative location to the forming star $\mathbf{r}_{\rm p,i}$ and relative velocity $\mathbf{v}_{\rm p,i}$.
\begin{equation}
    M=\Sigma_i \mathbf{m}_{\rm p,i}
\end{equation}
is the total mass of all accreting particles.
We plot $J$ over the total particle mass that falls onto the star. 
Note that the total particle mass is a factor of 2 higher than the mass of the stars particles because of the efficiency factor of $0.5$ in the sink accretion recipe to account for loss via outflows that are at best poorly resolved in these simulations.
For particles that have already accreted in the brief time span before the first snapshot after star formation, we set their angular momentum to 0.

The plot shows scatter of several orders of magnitude in specific angular momentum ranging from $\sim10^{-5}$ km s$^{-1}$pc for one very low-mass star up to $\sim10^0$ km s$^{-1}$pc.
There is especially wide scatter in the distribution of specific angular momentum for stars in the range of $\sim 0.1$ M$\odot$ to $\sim 1$ M$\odot$.
Despite the scatter, the plot also reveals a trend of increasing specific angular momentum with increasing stellar mass.
The trend is clearly visible in the arithmetic mean value per mass bin and it is even more pronounced when looking at the median and geometric mean per mass bin. 
This correlation can be understood when considering the possibility and probability of infall.
While the stars with lowest mass in the simulation accrete almost all of their material during the initial collapse phase and thereby from a rather narrow spatial reservoir, the average amount of late infall increases with stellar mass.
The trend of increasing specific angular momentum for increasing stellar mass is therefore a direct consequence of late infall. 

\begin{figure}
\centering
\includegraphics[width=\linewidth]{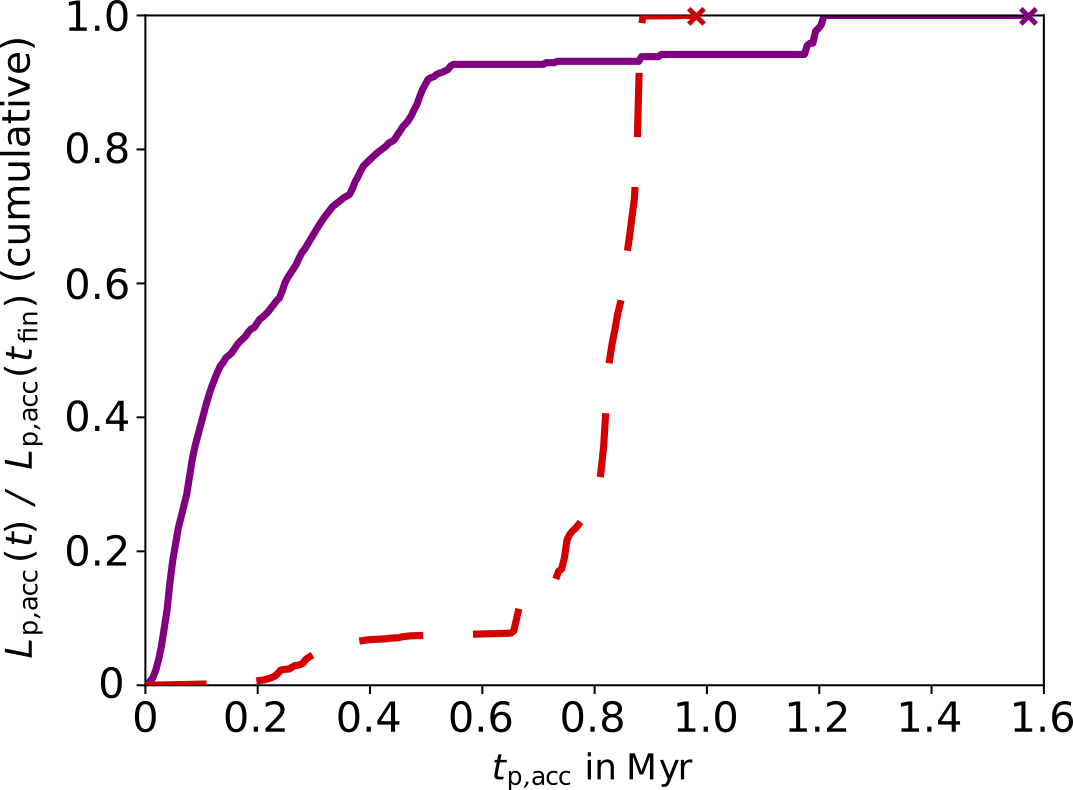}
\caption{Cumulative angular momentum of accreting tracer particles for star A and star B measured at the first snapshot after star formation over the time when the corresponding tracer particles accrete onto the star. The angular momentum is plotted relative to the total angular momentum of the accreting particles. The violet solid line shows the cumulative angular momentum for star A, the red dashed line for star B. The crosses mark the age of the stars at the end of the simulation.}\label{fig:t_L}
\end{figure}

Comparing the two stars selected for the detailed analysis of tracer particles shown in \Fig{tracer_cum_46_123} and highlighted in \Fig{M_J_M_Rc} illustrates the correlation of higher angular momentum and late infall.   
Star A accretes its mass from a relatively narrow reservoir of the prestellar core.
Late infall is practically absent for this star as seen in the left panels of \Fig{tracer_cum_46_123}. 
In contrast, star B undergoes substantial late infall and accretes its mass from a spatially more extended reservoir as demonstrated in the right panels of \Fig{tracer_cum_46_123}. 
As (specific) angular momentum increases with the distance from the source, 
material that is falling in late has substantially higher (specific) angular momentum. 
This can be clearly seen in \Fig{t_L}, where we plot the angular momentum of all accreting tracer particles over the time when they accrete relative to their total angular momentum. 
For star B, the angular momentum budget of particles accreting after 600 kyr contributes to more than $90 \%$ of the total angular momentum.
Even in the case of star A that only experiences almost negligible late infall in terms of mass, this tiny amount of infall contributes to almost $10 \%$ of the total angular momentum of accreting particles.

Moreover, the relative velocities with respect to the source are likely higher for gas that is not gravitationally bound to the collapsing progenitor core compared to gas that is bound. This further increases the angular momentum budget of late accretors.
Considering that we can account for material from possible late infall in our calculation, it is neither a surprise to see wide scatter of specific angular momentum for the individual stars nor to find high values of specific angular momentum that exceed observational estimates derived from dense cores as the accreting reservoir \citep[e.g.][]{Pineda+2019}.

\subsubsection{Centrifugal radii of accreting tracer particles}
The obtained estimates of the angular momentum budget can help us to gain insight in the disk formation process even though disks are not resolved in this study.
Given the specific angular momentum reservoir of each star,
we estimate the centrifugal radius
\begin{equation}
    R_{\rm c} = J^2/(G M),
\end{equation}
 where $G$ is the gravitational constant. 
 The centrifugal radius is the radius where gravitational and centrifugal force are in balance.
In the right panel of \Fig{M_J_M_Rc}, we plot the centrifugal radius over total mass of accreting particles. Apart from one outlier with $R_c<1$ au, which is a star that gained almost all of its mass at the very beginning of its formation, the other objects have generally large centrifugal radii in comparison to observed disk sizes.  
While some systems have radii $R_{\rm c} < 100$ au, there are more than a dozen (i.e., more than 10 \% of the sample) with $R_{\rm c} > 1$ pc, some even show $R_{\rm c}$ of hundreds of parsecs. 
Such large disk radii are of course unrealistic because we know from observations that disks often are only about $100$ AU or less in size. 
The very large values of $R_{\rm c}$ emphasize the importance of angular momentum transport in the star and disk formation process.
The large scatter of $R_{\rm c}$ also shows the
short-comings of considering the protostellar envelope as the sole angular momentum reservoir for disk formation.

The data also reveal a subtle trend of increasing $R_{\rm c}$ with increasing stellar mass.
This trend is not reflected in the arithmetic mean values per mass bin, where there is an increases of $R_{\rm c}$ up to $\approx 1$ M$_{\odot}$ and a decrease for higher masses again. 
However, it is evident for the geometric mean of $R_{\rm c}$. 
The median value fluctuates more around $R_{\rm c} \approx 10^4$ au for total particle masses up to $\approx 1$ M$_{\odot}$ (corresponding to stellar mass of $\approx 0.5$ M$_{\odot}$), but there is a mild increase of $R_{\rm c}$ toward total particle masses higher than $1$ M$_{\odot}$. 
We also expect a more visible trend in simulations of a larger cloud, where the stars evolve for a longer time because of the increasing probability of late infall for higher mass stars.

Such a trend of increasing disk radii with increasing stellar mass such as seen in the geometric mean and median is consistent with observations \citep{Long+2022}.
It is difficult to draw a conclusion about the disk properties without resolving the actual disk formation process of course, but the analysis demonstrates that stars that undergo late infall are fed with material that initially has very high specific angular momentum. 
A lot of the angular momentum must be transported away from the gas such that it can approach the star.
Nevertheless, the analysis strongly suggests that the probability of (re-)forming larger disks for late accretors is supposedly higher than for primordial disks that solely form from collapse of the progenitor core. 
We also point out that we expect more scatter of disk sizes for low-mass stars in the range of $\approx0.1$ M$_{\odot}$ to $\approx1$ M$_{\odot}$ compared to stars above $1$ M$_{\odot}$, where we generally expect larger disk sizes as they are more prone to late infall (see \Fig{sink_mass}).

Unless the angular momentum is efficiently transported away from the infalling material, the primordial disk can grow significantly in size or even be surrounded by a large, possibly misaligned second-generation disk such as seen in simplified parameter studies \citep{Thies+2011, Kuffmeier+2021}.
The increased probability of late infall for higher stellar masses opens an intriguing explanation of the stellar mass-disk size relation \citep{Andrews+2018,Long+2022}.
Considering that material from late infall has high angular momentum, it implies that on average disks become larger for increasing stellar mass, which is consistent with observations.
In addition, the statistical manner of late infall also explains the significant scatter of disk sizes at later stages. 
We suggest that evolved stars with small disks are the stars that did not experience substantial late infall, while the ones with large disks are the ones that experienced late infall during the last $\sim 100 000$ to 1 million years. 

The models strongly suggest that disk sizes will be larger for late accretors as they have a larger angular momentum budget.
The presented models do not resolve the disk, thus we cannot test the hypothesis here. 
Upcoming models with high enough resolution to resolve disk formation from stellar birth throughout the simulation including late infall, however, will allow us to answer to what extent the substantially larger angular momentum budget of late accretors is reflected in their disk sizes.

Our analysis of tracer particles shows in detail that gas can initially be more than 1 pc away, but yet accrete onto the protostar about 1 Myr later.
This revised picture with an optional second phase has several profound consequences on our understanding of star formation.
Hydrodynamical simulations of cloudlet captures already demonstrated that late infall can lead to the formation of a second-generation disk \citep{Kuffmeier+2020} and even to systems with misaligned inner and outer disks \citep{Kuffmeier+2021} that are an intriguing explanation of shadows detected in scattered light observations \cite{Ginski+2021}.
In the remaining part of the paper, we focus on how late infall affects the observed bolometric temperature and thereby the classification of the evolutionary stage of a YSO.

\begin{figure*}
\centering
\includegraphics[width=\textwidth]{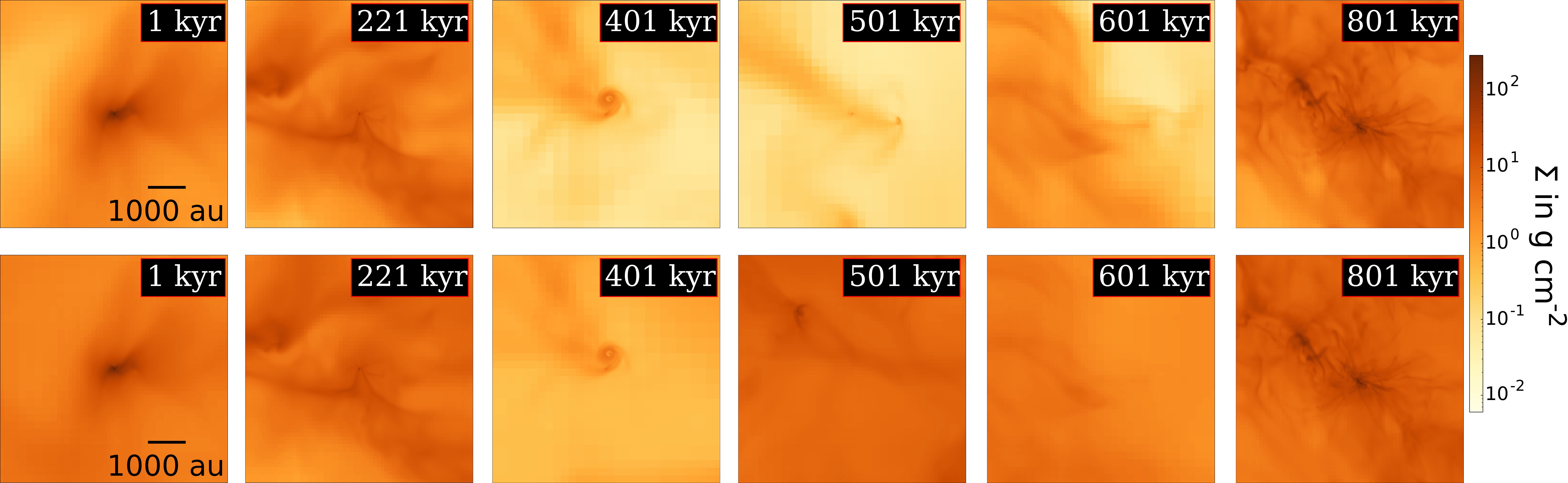}
\caption{From left to right: column densities $\Sigma$ in an area of $\pm 3000$ au around the protostar at $t=1 \unit{kyr}$, $t=221 \unit{kyr}$, $t=401 \unit{kyr}$, $t=501 \unit{kyr}$, $t=601 \unit{kyr}$ and $t=801 \unit{kyr}$. In all snapshots, the staris located at the center of the image.
Top panels: $\Sigma$ computed for a vertical extent of the column of $\Delta z=\pm 3000$ au.
Bottom panels: $\Sigma$ for the whole vertical extent of the box of 4 pc.
). }\label{fig:Sigma_maps}
\end{figure*}

\subsection{Late accretion as a source of (apparent) rejuvenation of YSOs}
\begin{figure*}
\centering
\includegraphics[width=\linewidth]{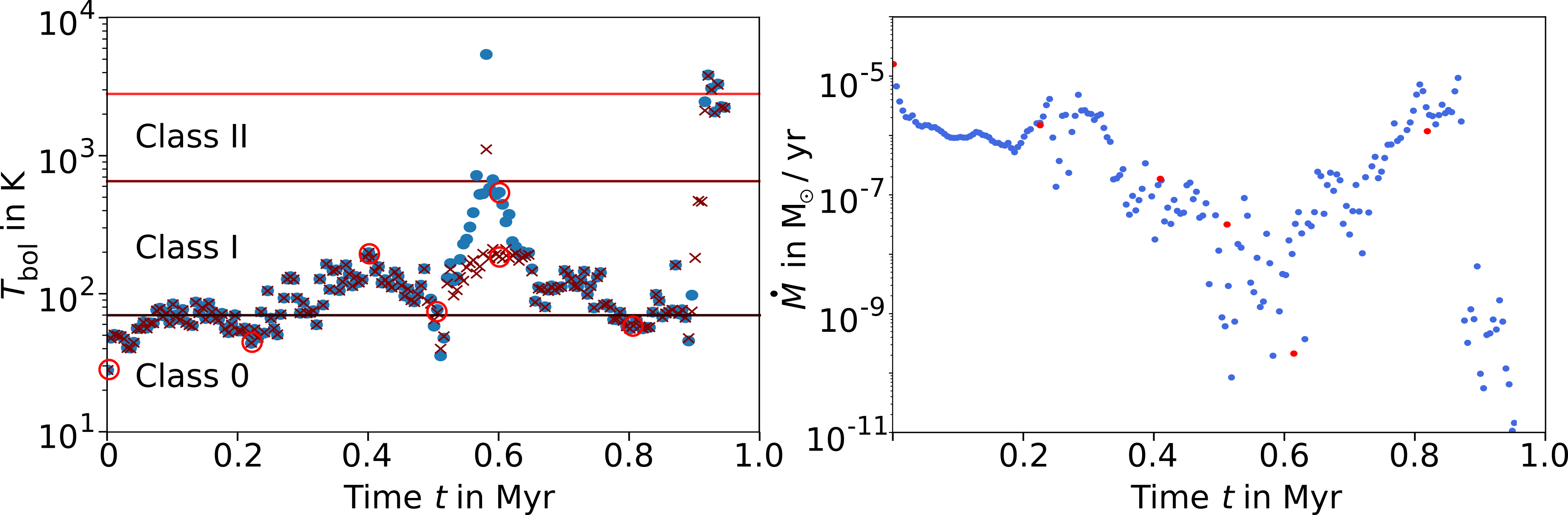}
\caption{Evolution of bolometric temperature $T_{\rm bol}$ (left) and $\dot{M}_{\rm acc}$ (right). The horizontal lines in the left panel mark the distinction between Class 0, Class I and Class II. Left panel: the blue dots mark the uncorrected values, the red crosses mark the values corrected for the background emission by comparing the flux computed at aperture sizes of 50000 AU and 3000 AU.} The red circles (left) and red dots (right) mark the snapshots for which the column densities are shown in Fig. 4.\label{fig:Tbol}
\end{figure*}

Protostars are commonly categorized according to a class system ranging from Class 0, when the protostar is deeply embedded to Class II/III, when the envelope has fallen onto the protostellar system and the protostar and its disk are not embedded anymore. 
The classification is commonly used as an observational tracer of the evolutionary stage of the protostar with increasing Class corresponding to a later evolutionary stage. 
In other words, the less embedded the protostar is, the higher the class it is classified as, and hence the later the presumed evolutionary stage of the star. 

However, we demonstrated that protostars move through the cloud after their formation and that some protostars enter regions of high densities, in which they are fed with fresh material later on. 
In \Figure{Sigma_maps}, we show column density maps from star B at different evolutionary stages within $\pm 3000$ au from the star (upper panel) and for the whole depth of the box (lower panel).
The column density maps illustrate that the protostar becomes less embedded a few 100 000 years after its initial formation phase, but enters a region of higher density again at about 800 kyr after its formation. 
Our hypothesis is therefore that the protostar is classified with a lower Class (and thereby an earlier evolutionary stage) during/after the infall event than prior to it. 
To test the hypothesis, we selected the star that experienced late infall events and post-processed its environment with the radiative transfer code \radmc \citep{Dullemond2012} at various times after star formation. 

This allows us to produce spectral energy distributions (SEDs) for the selected stars before during and after an infall event. 
As the resolution of the underlying MHD models is not sufficiently high to resolve the circumstellar disk, we do not use the slope of the SED for classification, but apply the criterion of bolometric temperature to diagnose the class of a YSO. 
A YSO is classified as Class 0 for temperatures below 70 K, as Class I for temperatures in the range 70 K to 650 K and as Class II in the range 650 K to 2800 K. 

The \radmc~simulation follows the methodology introduced in \cite{Frimann2016} and \cite{Jensen+2021}. In brief, the protostar is treated as a point source with an effective temperature either fixed at 2000~K or interpolated from pre-main sequence stellar evolution models, depending on the age of the protostar. The luminosity of the protostar is the sum of the accretion luminosity and the pre-main sequence luminosity, again interpolated from stellar evolution models.
When computing the bolometric temperature, we compute both a raw and a background-corrected temperature. In the later case, we compute the flux from the protostar in a larger aperture (50,000~au instead of 3,000~au) and determine the background flux from the difference in flux between the two apertures. Uncorrected temperature are marked as blue dots in Fig. \ref{fig:Tbol}, while red crosses indicate the background-corrected temperatures. 
Only minor difference of $\lt$ 10\% are seen, except around 0.5--0.7~Myr, when several nearby protostars enter the larger aperture.

We find two episodes during which the bolometric temperatures drops to values below the threshold for Class 0 objects of 70 K.
The first episode at $\approx 500$ kyr is mostly caused by an increase of the background emission, while the protostar does not become significantly more embedded again. 
This can be seen when comparing the column densities in \Fig{Sigma_maps} within a vertical extent of $\pm3000$ au (upper panels) to the column densities measured along the entire column of the box (lower panels).  
Also the accretion rate (right panel in \Fig{Tbol}) does not increase during that period and remains low. 
Therefore, the drop at around 500 kyr is more visible in the uncorrected bolometric temperatures. Nevertheless, the YSO still appears as a YSO Class 0 object even when using the commonly applied observational method of accounting for the background correction, hence the correction is insufficient to filter out the background emission completely. 

However, we also find that the protostar is classified as Class I before the infall event at $\approx 800$ kyr, but classified as Class 0 during/after the event. 
At this stage, the protostar enters a dense region and becomes more embedded again as clearly seen in \Fig{Sigma_maps}. 
As shown in the right panel in \Fig{Tbol}, the accretion rate increases from as little as $M_{\rm acc}\sim10^{-10}$ M$_{\odot}$ yr$^{-1}$ to $M_{\rm acc}>10^{-5}$ M$_{\odot}$ yr$^{-1}$ during the phase when the protostar becomes more embedded again. 
As a result, the uncorrected and the corrected bolometric temperatures are approximately the same in this range. 
They both drop to values below the threshold value of 70 K.
Fluctuations in the bolometric temperature due to variations in the accretion rate during the first $\sim1$ Myr are also in agreement with earlier findings by \cite{WuchterlKlessen2001}.
The results confirm our hypothesis that the protostar appears as a younger object during and after infall when using the Class criteria of bolometric temperature as a tracer of the evolutionary stage.

\section{Discussion}\label{sec12}
While late infall events onto an already existing star-disk system were studied in specifically designed parameter studies of captured cloudlets, it was unclear how likely or frequent such events are.   
Our simulations show that accretion events after the initial collapse phase happen and that they can even provide a substantial amount of the mass budget of the star-disk system. 
In these simulations, we do not adequately resolve the disk, but the amount of mass that the star is fed with during late accretion events can exceed measured disk masses by orders of magnitude. This implies that the disk can be refreshed or even entirely rejuvenated with plenty of material that is from a different location in the cloud and thereby likely has a different chemical composition. 
The simulations do not only show that late accretion events during which the star is fed with fresh material from a different reservoir \textit{can} occur, they also show that late accretion events where more than 0.1 solar mass accrete onto the protostar \textit{are the norm} rather than the exception for solar mass stars. 

Our results demonstrate that protostar formation often is a two-phase process. 
Initially, the star forms as a consequence of gravitational collapse of a prestellar core. 
This phase is analogous to the classical paradigm of star formation. 
In fact, we find cases where the final protostellar mass at the end of the simulation is
almost entirely determined by the collapse phase of the progenitor core.

However, in many cases the protostellar mass is not finally set after this phase.
Many protostars in our simulation only have relatively small mass after the initial collapse phase with respect to their mass at the end of the simulation. 
The reason for this is that protostars may undergo substantial accretion and accumulate more mass even when the accretion rate has already stalled beforehand. 
Our analysis demonstrates that the probability of this second phase, the late accretion phase, increases with increasing final mass of the star. 
As first shown by \cite{Pelkonen+2021}, especially for higher mass stars, the final distribution of stellar masses is not determined by the collapse of the progenitor core. 
Even for solar-mass stars, \textit{on average} about $50 \%$ of the finally accreted mass is initially not gravitationally bound to the collapsing progenitor core. 
In addition, an independent study using Lagrangian tracer particles with \enzo\ \citep{Collins+2023} also supports the result of diverse trajectories of accreting gas onto a protostar in a turbulent molecular cloud environment.

At the current stage, observations of growing disks are commonly interpreted in a framework of isolated disks. 
In this framework, larger disks especially at later stages are potentially interpreted as a sign of viscous spreading and a violation of MHD wind models \citep[][]{Trapman+2020,Manara+2022}. 
Our results, however, strongly suggest that large disks are in reality the result of a star-disk system that underwent late accretion of high angular momentum material. 
This implies that it is inadequate to interpret the evolution of protostellar disks solely as isolated systems that are detached from the protostellar environment. 
Considering the high angular momentum of infalling material, it is not even enough to constrain the region of infall to the envelope.
It is therefore expected to detect both small and large disks as a result of the statistical scatter of infall without the necessity of imposing long-lasting high viscosity parameters for accretion disks.

Our results show the importance of late infall as a mechanism that contributes a substantial amount of material to the system. 
At first glance, one might think that the model results are at tension with observations because visible tracers of late infall appear to be rare. However, there are multiple aspects that explain that both, model results and observations, are in agreement with each other \citep[see also discussion in][]{Gupta+2023}.
First, a protostar can become very embedded again during a late accretion event. Our synthetic observations demonstrate that using bolometric temperature as a tracer, a rejuvenated protostar would be classified as a Class 0 or Class I object, and thereby be miscategorized as a much younger object than it really is. 
Additional observational criteria of classifying the evolutionary stage might help to eventually distinguish between a "true" Class 0 and a rejuvenated Class 0 object. 
For instance, we find that the object would only be classified as borderline Class 0/I during the rejuvenated stage, when applying $L_{\rm bol}/L_{\rm smm}$ as a tracer \citep{Andre+1993}. 
Nevertheless, the fundamental issue remains. Infall enhances the embeddedness of the protostar, and hence, observationally, the protostar appears younger than prior to the infall event. 

Second, late accretion events last for $\sim1$ to $\sim 100$ kyr. On the one hand, this means that it is possible to find objects with visible signs of late infall such as streamers considering protostellar evolution on timescales of several million years, on the other hand, it means that it is much more likely to observe a protostar that underwent late accretion without visible tracers of the actual infall event.
Third, streamers are relatively faint compared to the bright protostar and its disk and also become fainter again when material is transferred from the cloud to the protostar in form of streamers. Therefore, we expect a relatively low detection rate of streamers, even though the ratio of stars undergoing significant infall via streamers at some point during their formation and evolution is high.

\section{Conclusion}\label{sec13}

Carrying out MHD simulations of an evolving clump of ($4$ pc)$^3$ with the adaptive mesh-refinement code \ramses, we find that (late) infall scales with final stellar mass \citep[such as previously demonstrated][]{Pelkonen+2021}. 
As part of the heterogeneous formation process, stars with similar final mass can have very diverse accretion reservoirs. 
We demonstrate that late infall can feed the star with additional material. 
The amount of infalling material is small for very low mass stars, but their relative contribution to the final mass increases with increasing stellar mass $M_*$. 
Given that observed disk masses are only $\sim 10^{-2} M_{*}$, even a small fraction of infalling mass can still be a substantial amount of the disk mass.
Infall-induced structures are therefore a very intriguing explanation to trigger planet formation, when considering planet formation as part of the larger picture of star formation.

Moreover, the possibility of infall implies a huge scatter in the angular momentum budget of individual protostars.
Our analysis also emphasizes the importance of angular momentum transport during the formation process of disk to regulate their sizes. 
The large scatter together with the large values of angular momentum show the limitations of considering only a collapsing envelope as the source of "infall". 
Stars that undergo late accretion are fed with material that was located far beyond the progenitor core, hence it has very high angular momentum.
As a result, stars experiencing late infall accrete their mass from a reservoir with higher specific angular momentum.
The increasing probability of late infall for increasing final stellar mass implies a mild correlation of centrifugal radius and stellar mass.
The trend of increasing centrifugal radius with increasing stellar mass is very subtle and requires follow-up investigations at higher resolution. If the trend is indeed inherited at disk scales, it would be in line with observations of increasing disk sizes for increasing mass of the disk-hosting star \citep[see Figure 5 in][]{Long+2022}.
The statistical nature of late infall including the substantial scatter of the angular momentum budget can also explain the observed scatter in disk sizes.
Our analysis strongly suggests that stars with extraordinarily large disks such as GO Tau were prone to substantial late infall with high angular momentum, while stars with small disks experienced at most modest infall and/or have evolved for long enough since the last infall event to allow the disk to shrink efficiently again. 
Future models resolving the disks during and after infall events will clarify to what extent the angular momentum of disks is inherited from infalling material that stems from locations far beyond the collapsing prestellar core. 
We highlight, however, that interpreting observations in the framework of isolated disks instead of taking into account for the possibility of post-collapse infall yields a false picture of the disk's history.

Last, but not least, we investigated the result of late infall on the classification of the protostar. 
The protostar becomes temporarily more embedded again when the star enters a region of higher density. 
Radiative transfer simulations reveal that as a consequence the protostar would be classified as a Class 0 object despite the fact that it is already $\sim$1 million year old. 
In short, infall acts as a rejuvenation cure for the protostar and its disk.

\section{Data Availability Statement}
The datasets generated and/or analyzed during the current study are not publicly available due to ongoing research of other aspects of the datasets but are available from the corresponding author on reasonable request.

\section*{Acknowledgements}
We thank the anonymous referee for the positive feedback and a very constructive report that improved the quality of this manuscript. 
We also thank Jaime Pineda, Paola Caselli, Anna Miotello, Aashish Gupta, Jon Ramsey, Zhi-Yun Li, {\AA}ke Nordlund and Anders Johansen for helpful discussions leading to this work. 
MK has received funding for this project from the European Union's
Horizon 2020 research and innovation programme under the Marie Sk{\l}odowska-Curie grant agreement No.\~ 897524.
SJ acknowledges the financial support of the Max Planck Society.
TH acknowledges funding from the Independent Research Fund Denmark through grant No. DFF 8021-00350B.
%\begin{appendices}




%\end{appendices}


\bibliography{bibliography}
\bibliographystyle{unsrt}


\end{document}